\documentclass{redaction}%
\usepackage{nota}%
\usepackage{meta}%
\date{April 19th, 2016}%
\def\seqAAAaad{}%
\def\InputFigPaar{\includeStandaloneFig{1}{\figBBOXaar}}%
\def\seqAAAabb{\Mpar \vspace{\Sssaut}\hspace{\Alinea}}%
\def\seqDDDacb{\bseqHHHabg{\abbrevDef 2.7}}%
\def\seqBBBabz{\bseqHHHaaq{\abbrevSectionB 2}}%
\def\MprefixNNNaaa{\Mprefixe{ova}{Projective Limits of State Spaces:\\[2pt] Quantum Field Theory without a Vacuum}}%
\def\seqDDDabf{\bseqHHHaat{\abbrevDefs 2.1 and 2.2}}%
\def\seqAAAacz{}%
\def\seqAAAaaw{\MStopList \Mpar \MList{4}}%
\def\seqAAAacl#1{\hypertarget{SECacl}{}\MsectionB{acl}{4.1}{#1}}%
\def\seqBBBabp{\bseqHHHaat{\abbrevTheorem 2.9}}%
\def\seqAAAaaf{\vspace{\PsectionA}\Mpar \vspace{\Sssaut}\hspace{\Alinea}}%
\def\seqDDDacn{\bseqHHHabh{\abbrevSectionC 3.1}}%
\def\seqAAAack{}%
\def\seqBBBabd{\bseqHHHabh{\abbrevDef 2.11}}%
\def\seqAAAaax{\MStopList \Mpar }%
\def\seqDDDabh{\bseqHHHaat{\abbrevTheorem 3.9 and \abbrevProposition 3.11}}%
\def\seqAAAaas{\Mpar \MList{1}}%
\def\seqAAAabm#1{\hypertarget{SECabm}{}\MsectionA{abm}{3}{#1}}%
\def\seqAAAacw{\vspace{\PsectionA}\Mpar \vspace{\Sssaut}\hspace{\Alinea}}%
\def\seqBBBacu{\bseqHHHabi{\abbrevSectionC 3.2}}%
\def\seqAAAach{\Mpar \vspace{\Sssaut}\hspace{\Alinea}}%
\def\seqAAAacm{\vspace{\PsectionB}\Mpar \vspace{\Sssaut}\hspace{\Alinea}}%
\def\seqAAAacs#1{\hypertarget{SECacs}{}\MsectionB{acs}{4.2}{#1}}%
\def\seqBBBaap{\bseqHHHabh{\abbrevSectionC 2.1}}%
\def\seqBBBabi{\bseqHHHaat{\abbrevSectionC 3.1}}%
\def\seqAAAabl{}%
\def\seqAAAacf{\MStopList \Mpar }%
\def\seqAAAaaz{\Myfigure{2.2}{\hypertarget{PARaaz}{}}}%
\def\seqBBBaaz{\abbrevFigure \hyperlink{PARaaz}{2.2}}%
\def\InputFigPaba{\includeStandaloneFig{2}{\figBBOXaba}}%
\def\seqAAAacc{\Mpar \MList{1}}%
\def\seqBBBabg{\bseqHHHaat{\abbrevSectionB 3}}%
\def\seqAAAact{\vspace{\PsectionB}\Mpar \vspace{\Sssaut}\hspace{\Alinea}}%
\def\seqBBBabw{\bseqHHHabg{\abbrevProposition 2.1}}%
\def\seqAAAaak#1{\hypertarget{SECaak}{}\MsectionA{aak}{4}{#1}}%
\def\seqBBBaak{\abbrevSectionB \hyperlink{SECaak}{4}}%
\def\seqDDDaak{(\abbrevSectionB \hyperlink{SECaak}{4})}%
\def\seqBBBacr{\bseqHHHabh{\abbrevSectionC 3.2}}%
\def\seqDDDacr{\bseqHHHabh{\abbrevSectionC 3.2}}%
\def\seqDDDace{\bseqHHHabg{\abbrevFigure 2.1}}%
\def\seqBBBaav{\bseqHHHaat{\abbrevSectionB 1}}%
\def\seqAAAaam#1{\hypertarget{SECaam}{}\MsectionA{aam}{2}{#1}}%
\def\seqBBBaam{\abbrevSectionB \hyperlink{SECaam}{2}}%
\def\seqAAAabx#1{\hypertarget{SECabx}{}\MsectionB{abx}{3.3}{#1}}%
\def\seqDDDaca{\bseqHHHabg{\abbrevSectionC 2.2}}%
\def\seqDDDacj{\bseqHHHabg{\abbrevProposition 2.9}}%
\def\seqAAAacv{}%
\def\seqAAAabe{\vspace{\PsectionB}\Mpar \vspace{\Sssaut}\hspace{\Alinea}}%
\def\seqAAAaaj#1{\hypertarget{SECaaj}{}\MsectionB{aaj}{3.1}{#1}}%
\def\seqDDDaaj{(\abbrevSectionC \hyperlink{SECaaj}{3.1})}%
\def\seqAAAaal{}%
\def\seqBBBaay{\bseqHHHabh{\abbrevDef 2.6 and \abbrevProposition 2.7}}%
\def\seqDDDabv{\bseqHHHaat{\abbrevProposition 2.6}}%
\def\seqAAAaco{\Myfigure{4.1}{\hypertarget{PARaco}{}}}%
\def\seqBBBaco{\abbrevFigure \hyperlink{PARaco}{4.1}}%
\def\seqDDDabu{\bseqHHHabg{\abbrevSectionC 2.1}}%
\def\seqDDDaai{(\abbrevSectionCs \hyperlink{SECabs}{3.2} and \hyperlink{SECabx}{3.3})}%
\def\seqBBBabq{\bseqHHHabi{\abbrevProposition 3.17} and \bseqHHHaaq{\abbrevTheorem 3.20}}%
\def\seqAAAaag#1{\hypertarget{SECaag}{}\MsectionA{aag}{5}{#1}}%
\def\seqBBBaag{\abbrevSectionB \hyperlink{SECaag}{5}}%
\def\seqAAAaac{\Mpar }%
\def\seqBBBabj{\bseqHHHaat{\abbrevSectionC 3.2}}%
\def\seqAAAaaq{\Myfigure{2.1}{\hypertarget{PARaaq}{}}}%
\def\seqBBBaaq{\abbrevFigure \hyperlink{PARaaq}{2.1}}%
\def\seqDDDabr{\bseqHHHaat{\abbrevTheorem 2.11}}%
\def\seqAAAaby{\vspace{\PsectionB}\Mpar \vspace{\Sssaut}\hspace{\Alinea}}%
\def\seqAAAaab{\Mpar }%
\def\seqAAAaao{\vspace{\PsectionB}\Mpar \vspace{\Sssaut}\hspace{\Alinea}}%
\def\seqAAAacy{\Acknowledge }%
\def\seqDDDaci{\bseqHHHabg{\abbrevTheorem 2.8}}%
\def\seqAAAabs#1{\hypertarget{SECabs}{}\MsectionB{abs}{3.2}{#1}}%
\def\seqBBBabs{\abbrevSectionC \hyperlink{SECabs}{3.2}}%
\def\seqBBBabk{\bseqHHHaat{\abbrevSectionB 4}}%
\def\seqDDDabo{\bseqHHHaat{\abbrevProposition 2.8}}%
\def\seqAAAaat{\MStopList \Mpar \MList{2}}%
\def\seqAAAacx{\Mpar }%
\def\seqAAAaae#1{\hypertarget{SECaae}{}\MsectionA{aae}{1}{#1}}%
\def\seqBBBaae{\abbrevSectionB \hyperlink{SECaae}{1}}%
\def\seqAAAabt{\vspace{\PsectionB}\Mpar \vspace{\Sssaut}\hspace{\Alinea}}%
\def\seqAAAabn{\vspace{\PsectionB}\Mpar \vspace{\Sssaut}\hspace{\Alinea}}%
\def\seqBBBabc{\bseqHHHabh{\abbrevProposition 2.10}}%
\def\InputFigPacp{\includeStandaloneFig{3}{\figBBOXacp}}%
\def\seqAAAaau{\MStopList \Mpar \MList{3}}%
\def\MsectionBiblio#1{\MsectionA{ada}{A}{#1}}%
\def\seqAAAaan#1{\hypertarget{SECaan}{}\MsectionB{aan}{2.1}{#1}}%
\def\seqBBBaan{\abbrevSectionC \hyperlink{SECaan}{2.1}}%
\def\seqAAAacq{\Mpar \vspace{\Sssaut}\hspace{\Alinea}}%
\def\seqBBBacg{\bseqHHHabg{\abbrevSectionB 3}}%
\def\seqAAAacd{\MStopList \Mpar \MList{2}}%
\def\seqAAAaah#1{\hypertarget{SECaah}{}\MsectionB{aah}{2.2}{#1}}%
\def\seqBBBaah{\abbrevSectionC \hyperlink{SECaah}{2.2}}%
\def\figBBOXaar{0.0pt 0.0pt 469.45116pt 213.38689pt}%
\def\figBBOXaba{31.53348pt 0.0pt 330.29248pt 199.17267pt}%
\def\figBBOXacp{20.71484pt 0.0pt 507.26169pt 213.39772pt}%
\def\bseqJJJabe{\cite{abn}}%
\def\bseqJJJaaa{\cite{aaw}}%
\def\bseqJJJaap{\cite{aaz}}%
\def\bseqJJJaav{\cite{abi}}%
\def\bseqJJJaab{\cite{aaf,aad,aae,aac}}%
\def\bseqJJJabf{\cite{aao}}%
\def\bseqJJJabd{\cite{aav,*aau}}%
\def\bseqJJJaah{\cite{abp}}%
\def\bseqHHHabi#1{\cite[#1]{aae}}%
\def\bseqJJJaax{\cite{aat}}%
\def\bseqJJJaau{\cite{aba}}%
\def\bseqJJJaai{\cite{abq,abo}}%
\def\bseqHHHabh#1{\cite[#1]{aaf}}%
\def\bseqJJJaam{\cite{aai,*aah}}%
\def\bseqJJJaac{\cite{aag}}%
\def\bseqHHHaac#1{\cite[#1]{aag}}%
\def\bseqJJJaar{\cite{aaw,aap}}%
\def\bseqJJJaan{\cite{abk,*abr}}%
\def\bseqJJJaaz{\cite{aaj,*aab}}%
\def\bseqHHHaag#1{\cite[#1]{abq}}%
\def\bseqJJJaal{\cite{aan,*aax}}%
\def\bseqHHHabg#1{\cite[#1]{aac}}%
\def\bseqJJJaaq{\cite{abl}}%
\def\bseqHHHaaq#1{\cite[#1]{abl}}%
\def\bseqJJJaak{\cite{abd,abg}}%
\def\bseqJJJaat{\cite{aad}}%
\def\bseqHHHaat#1{\cite[#1]{aad}}%
\def\bseqJJJabb{\cite{abe}}%
\def\bseqJJJabc{\cite{abf}}%
\def\bseqJJJaas{\cite{aap}}%
\def\bseqHHHaas#1{\cite[#1]{aap}}%
\def\bseqJJJaba{\cite{abh}}%
\def\bseqJJJaay{\cite{aaa,*aaq,*aar}}%
\def\bseqJJJaaf{\cite{aas,*abm}}%
\def\bseqJJJaad{\cite{aam,*aal,*aak}}%
\def\bseqJJJaaj{\cite{abb,abc,aap}}%
\def\bseqJJJaae{\cite{aay}}%
\def\bseqHHHaaw#1{\cite[#1]{abc}}%
\def\bseqJJJaao{\cite{abg,abj}}%
\begin{document}%
\normalsize\normalfont%
\clearpage\MprefixNNNaaa%
\authAffilAbstrac{\seqAAAaab Instead of formulating the states of a Quantum Field Theory (QFT) as density matrices over a single large Hilbert space, it has been proposed by Kijowski {\bseqJJJaaa} to construct them as consistent families of \emph{partial} density matrices, the latter being defined over small 'building block' Hilbert spaces.
In this picture, each small Hilbert space can be physically interpreted as extracting from the full theory specific degrees of freedom.
This allows to reduce the quantization of a classical field theory to the quantization of \emph{finite-dimensional} sub-systems, thus sidestepping some of the common \emph{ambiguities} (specifically, the issues revolving around the choice of a 'vacuum state'), while obtaining robust and well-controlled quantum states spaces.%
\seqAAAaac The present letter provides a self-contained introduction to this formalism, detailing its motivations as well as its relations to other approaches to QFT (such as conventional Fock-like Hilbert spaces, path-integral quantization, and the algebraic formulation). At the same time, it can serve as a reading guide to the series of more in-depth articles {\bseqJJJaab}.
\seqAAAaad%
}%
\Mtableofcontents%
\Mnomdefichier{ova1}%
\seqAAAaae{Motivation: Quantization Ambiguities in Quantum Field Theory}%
\seqAAAaaf Many choices have to be made in the quantization of a classical theory. Assuming one is following the canonical quantization path (see {\seqBBBaag} for further discussion of the relevance for path-integral approaches of the issues discussed here), the first step is to choose a complete set of basic variables for the theory. Heuristically, these are the variables for which the semi-classical limit will work best, hence their choice should ideally reflect the observables against which the classical theory of interest has been best tested and confirmed.%
\Mpar \vspace{\Sssaut}\hspace{\Alinea}The next step is to find a representation of these basic variables as operators on a suitable Hilbert space $\mathcal{H} $, namely a mapping $f \mapsto  \hat{f}$ such that \begin{equation}\label{quBrk}\big[ \hat{f},\,  \hat{g} \big] = i \,  \widehat{\left\{ f,\,  g \right\}}\end{equation} (where $\left[ \, \cdot \, , \, \cdot \,  \right]$ denotes the commutator of operators, while $\left\{ \, \cdot \, , \, \cdot \,  \right\}$ denotes the Poisson brackets of classical observables). At this point, quantum field theory (in a broad sense, namely quantum theories meant to encompass \emph{infinitely} many degrees of freedom) differs crucially from quantum mechanics (dealing with the quantum counterparts of classical systems that have \emph{finitely} many degrees of freedom).
The tools from geometric quantization {\bseqJJJaac} (that we will discuss further in {\seqBBBaah}) provide a clear and detailed understanding of the canonical quantization of finite dimensional systems, including a parametrization of available choices (aka.~quantization ambiguities).
In some cases, it may even turn out that there is no choice at all, because the Poisson-algebra of interest admits only one suitable representation: this is for example the content of the Stone-von-Neumann theorem {\bseqJJJaad} in the case of linear systems.%
\Mpar \vspace{\Sssaut}\hspace{\Alinea}By contrast, the representation theory for infinite dimensional system tends to be very involved. Even in the simplest case of a free scalar field on Minkowski spacetime, it is known that there exist infinitely many \emph{inequivalent} representations, and although it has been possible, in this very special case, to fully classify them {\bseqJJJaae}, this classification is so complex that it gives little insight on how to choose one.
As a way out, a pragmatic way of selecting a good representation among these too numerous options is to single out a distinguished quantum state, the \emph{vacuum}: it is indeed possible, via the so-called GNS construction {\bseqJJJaaf} to 'seed' a full representation $\mathcal{H} _{\Omega }$ from a single state $\Omega $ (to specify the latter, even before we are equipped with a Hilbert space, we can give the corresponding expectation values of all products of the basic variables, aka.~the $n$-point functions, see \bseqHHHaag{part III, def.~2.2.8}). This approach has established itself as the standard way to think about quantum field theory, at least in the context of Minkowski spacetime, where the vacuum may be selected by requiring it to be invariant under all spacetime symmetries (ie.~under the Poincaré group).%
\Mpar \vspace{\Sssaut}\hspace{\Alinea}However, one should keep in mind that the initial choice of vacuum is deeply imprinted in the thus obtained representation. The only quantum states that can be written as (pure or statistical) states on $\mathcal{H} _{\Omega }$ are those that barely differ from the vacuum: at most discrete quantum excitations on top of the state $\Omega $ are allowed. The set of all states living on the representation $\mathcal{H} _{\Omega }$ is referred to as the vacuum \emph{sector}, in acknowledgment of the fact that there are many more quantum states beyond it (falling out of it because they lie too far away from the chosen vacuum), among whose some may actually be interesting for specific purposes \bseqHHHaag{part V}. An implication of the relative smallness of the vacuum sector is that the vacuum state need to be closely tailored to the \emph{dynamics}: otherwise, the time evolution would immediately kick the states out of $\mathcal{H} _{\Omega }$ (a precise statement of this heuristic expectation is given, for Poincaré-invariant QFTs, by the Haag no-go theorem, {\bseqJJJaah}).%
\Mpar \vspace{\Sssaut}\hspace{\Alinea}A radical alternative, prompted by the lack of a natural vacuum in the case of quantum field theory on \emph{curved} spacetime, is to use as state space the \emph{whole} set of possible quantum states over the chosen basic observables (each such state being specified, as explained above, by the expectations values it prescribes for all products of observables). This approach can be followed in the context of Algebraic Quantum Field Theory (AQFT, {\bseqJJJaai}): by shifting the focus from a particle picture to the \emph{local} and \emph{causal} structure of the quantum theory, AQFT provides tools to discuss the properties of quantum fields in the absence of an underlying Hilbert space.
The aim of the present letter is to argue that a projective definition of quantum field theory, as was introduced by Jerzy Kijowski {\bseqJJJaaa} and further developed by Andrzej Okołów {\bseqJJJaaj}, can provide a middle way between the conventional vacuum-based approach and the full algebraic one, retaining a \emph{constructive} description of the quantum state space {\seqDDDaai} while keeping enough \emph{flexibility} to accommodate a wide class of quantum states {\seqDDDaaj} and to decouple the subsequent implementation of the dynamics from the initial building of the state space {\seqDDDaak}.%
\Mpar \vspace{\Sssaut}\hspace{\Alinea}The work summarized in the following sections (and developed in details in {\bseqJJJaab}) was notably motivated by the specific difficulties encountered when one tries to formulate \emph{background independent} quantum field theories, rather than theories on a (possibly curved) background spacetime (eg.~to quantize general relativity itself in a non-perturbative way {\bseqJJJaak}).
It turns out that for background independent gauge theories (at least those with \emph{compact} gauge group), there does exist a \emph{preferred} vacuum state, the \longAL vacuum {\bseqJJJaal}, which is uniquely selected precisely by the requirement of background independence {\bseqJJJaam}.
Unfortunately, this vacuum has some unwanted properties.
One of them is that it is an \emph{eigenstate} of the variable conjugate to the gauge field, rather than a \emph{coherent state} like the usual Fock vacuum. Since states in the vacuum sector cannot differ too much from the vacuum, this makes it difficult to find semi-classical states among them {\bseqJJJaan}.
Another problem is that the GNS representation built on this vacuum lives on a \emph{non-separable} Hilbert space.
This particular issue may or may not go away once we identify quantum states that only differ by a change of coordinates (depending on how precisely this identification is carried out, see {\bseqJJJaao}) but in any cases it can lead to technical difficulties {\bseqJJJaap}. Paradoxically, non-separable Hilbert spaces seem too \emph{small}, because their orthonormal basis need uncountably many basis vectors, making it tempting to consider uncountable linear combinations while only countable ones are allowed: in other words, it is in this case even more likely that physically interesting states will lie out of the vacuum sector.
In an effort to overcome these difficulties, the projective quantization techniques that we will review in the next section have been applied to this kind of theories, resulting in a quantum state space that may have applications to the study of the semi-classical and cosmological sectors of quantum gravity {\bseqJJJaaq}.
\seqAAAaal%
\Mnomdefichier{ova2}%
\seqAAAaam{Systematic Quantization of Infinite-dimensional Systems}%
\seqAAAaan{Building an infinite-dimensional theory from a collection of partial descriptions}%
\seqAAAaao The key observation underlying Kijowski's projective formalism {\bseqJJJaar} is that a given experiment can only measure finitely many observables. Thus, we never need to consider at once the full, \emph{infinite}-dimensional phase space $\mathcal{M} _{\infty }$ of a field theory: it is sufficient to work in a small, partial phase space $\mathcal{M} _{\eta }$ that extracts from $\mathcal{M} _{\infty }$ just the degrees of freedom (dof.\footnote{By a dof.\ we mean a \emph{pair} of conjugate variables.}) relevant for the experiment at hand (throughout the present letter, the symbol $\eta $ will be used to denote a selection of finitely many dof.~out of the full theory, and we will call $\eta $ a \emph{label}).%
\Mpar \vspace{\Sssaut}\hspace{\Alinea}In order to use such a collection of finite-dimensional partial phase spaces $\big( \mathcal{M} _{\eta } \big)_{\eta }$ to completely specify a field theory, we need to ensure that the different partial theories are consistent with each other {\seqBBBaap}:%
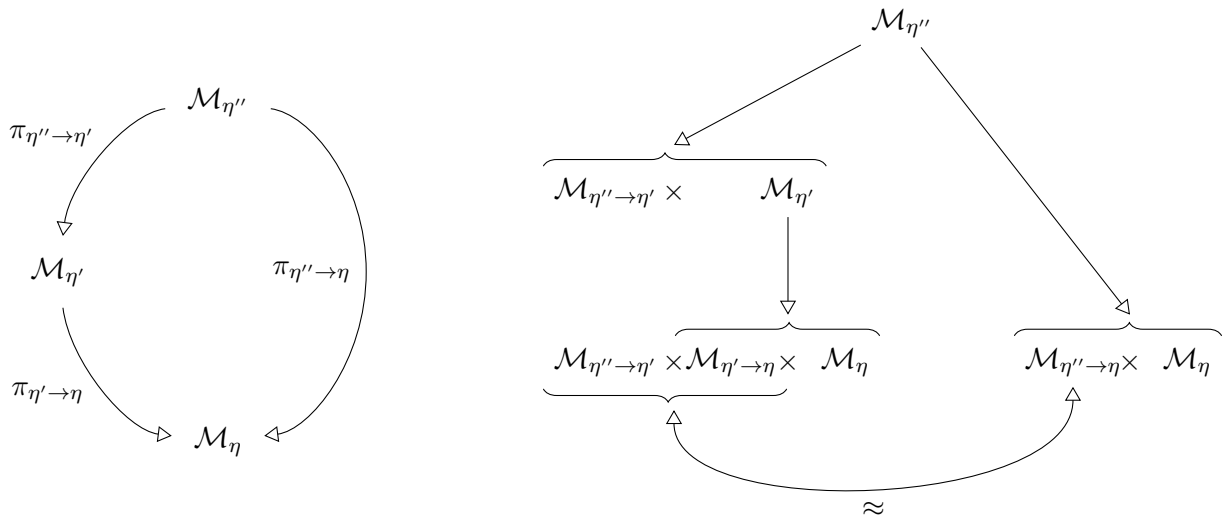
\seqAAAaaq {Three-spaces consistency for projective systems (left side), reformulated in terms of factorizations (right side)}{{\InputFigPaar}}%
\seqAAAaas first, we need a way to express the relations between the dof.\ in different labels. We will write $\eta  \preccurlyeq  \eta '$ if all dof.~contained in $\eta $ are also contained in $\eta '$ (we will also say that $\eta $ is \emph{coarser} as $\eta '$, or that $\eta '$ is \emph{finer} as $\eta $). This means that any observable $f_{\eta }$ on $\mathcal{M} _{\eta }$ corresponds to an observable $f_{\eta '}$ on $\mathcal{M} _{\eta '}$, and, by duality\footnote{To see that $\pi _{\eta '\rightarrow \eta }$ is uniquely specified once we know the mapping $f_{\eta } \mapsto  f_{\eta '}$ between observables, one can consider a complete set of observables (aka.~coordinates) on $\mathcal{M} _{\eta }$}, that there exists a projection $\pi _{\eta '\rightarrow \eta }$ from $\mathcal{M} _{\eta '}$ to $\mathcal{M} _{\eta }$ such that $$f_{\eta '} = f_{\eta } \circ  \pi _{{\eta '\rightarrow \eta }}$$%
\seqAAAaat the predictions for a given experiment, as calculated in a partial theory $\eta $, should be independent of the choice of $\eta $ (provided $\eta $ is fine enough to hold all pertinent dof.). Thus, in particular, the Poisson brackets between two observables $f_{\eta }$ and $g_{\eta }$ on $\mathcal{M} _{\eta }$ should agree with the Poisson brackets between the corresponding observables on a finer $\mathcal{M} _{\eta '}$. Expressed in terms of the just introduced projection $\pi _{\eta '\rightarrow \eta }$, this reads \begin{equation}\label{sympProj}\left\{ f_{\eta } \circ  \pi _{{\eta '\rightarrow \eta }}, g_{\eta } \circ  \pi _{{\eta '\rightarrow \eta }} \right\} = \left\{ f_{\eta }, g_{\eta } \right\} \circ  \pi _{{\eta '\rightarrow \eta }}\end{equation}%
\seqAAAaau it should be possible to consider composite experiments made of two (or more) sub-experiments, each of which can be described within a different partial theory\footnote{We are discussing the classical theory here. The quantum theory is more subtle, since one could argue that, due to the principle of complementarity, some sub-experiments may be mutually excluding. However, the case for the directedness of the label set can still be made, see {\seqBBBaav}.}. In other words, for any $\eta ,\eta '$, there should exist $\eta ''$ such that $\eta ,\eta ' \preccurlyeq  \eta ''$. This property is called \emph{directedness} of the set of labels.%
\seqAAAaaw the relation between any two partial theories should be unambiguous. Thanks to the just mentioned directedness property, this can be ensured simply by requiring that the projections defined among three increasingly refined partial theories match as shown on the left part of {\seqBBBaaq}.%
\seqAAAaax In mathematical terms, this list of requirements can be summarized by saying that the collection $\big( \mathcal{M} _{\eta } \big)_{\eta }$ forms a projective (aka.~inverse) system and it ensures that $\mathcal{M} _{\infty }$ can be reconstructed from $\big( \mathcal{M} _{\eta } \big)_{\eta }$ (more precisely, a space $\mathcal{M} _{\text{lim}}$ can be constructed as the so-called projective limit of this system, that will, in general, be a distributional extension of $\mathcal{M} _{\infty }$, see {\seqBBBaay}).%
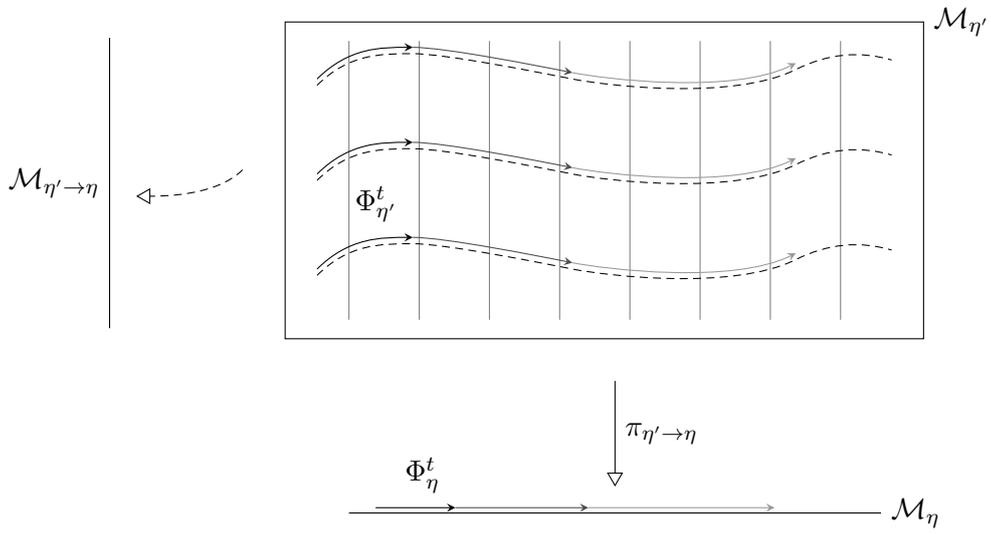
\seqAAAaaz {Factorization from a Poisson-brackets-preserving projection between phase spaces}{{\InputFigPaba}}%
\seqAAAabb A very important consequence of eq.~(\ref{sympProj}) is that it ensures that there exists a preferred \emph{factorization} of $\mathcal{M} _{\eta '}$ as $\mathcal{M} _{\eta '\rightarrow \eta } \times  \mathcal{M} _{\eta }$ (at least locally: there may be some topological obstructions preventing this factorization to hold globally). The way this factorization is obtained is as follow (see \bseqHHHaas{section 3.4} and {\seqBBBabc}). Any observable $f_{\eta }$ on $\mathcal{M} _{\eta }$ generates a Hamiltonian flow $\Phi ^{t}_{\eta }$ on $\mathcal{M} _{\eta }$, and $f_{\eta '}$ generates a corresponding Hamiltonian flow $\Phi ^{t}_{\eta '}$ on $\mathcal{M} _{\eta '}$. All the flows on $\mathcal{M} _{\eta '}$ generated in this manner from all possible observables on $\mathcal{M} _{\eta }$ together span joint orbits in $\mathcal{M} _{\eta '}$ (represented as dashed lines on {\seqBBBaaz}), and eq.~(\ref{sympProj}) ensures that any such orbit can (again, locally) be identified with $\mathcal{M} _{\eta }$. All what is left to do is then to define $\mathcal{M} _{\eta '\rightarrow \eta }$ as the corresponding quotient, ie.~the set of all such orbits (see {\seqBBBaaz}). $\mathcal{M} _{\eta '\rightarrow \eta }$ can thus be interpreted as holding dof.~that are complementary to the ones retained by $\mathcal{M} _{\eta }$ (since, by construction the dof.~in $\mathcal{M} _{\eta '\rightarrow \eta }$ Poisson-commute with the ones in $\mathcal{M} _{\eta }$).%
\Mpar \vspace{\Sssaut}\hspace{\Alinea}If we assume, for simplicity, that, for any $\eta  \preccurlyeq  \eta '$, the factorization $\mathcal{M} _{\eta '} \approx  \mathcal{M} _{\eta '\rightarrow \eta } \times  \mathcal{M} _{\eta }$ holds globally, we can translate the consistency condition illustrated on the left part of {\seqBBBaaq} into a condition written in terms of the factorizations {\seqBBBabd}, as shown on the right part of {\seqBBBaaq}. The meaning of this condition is that, when extracting from $\eta ''$ the dof.~corresponding to a much coarser partial theory $\eta $, the complementary dof.~that we discard should be the same whether we go from $\eta ''$ to $\eta $ in one step or in two, hence there should be a natural identification $\mathcal{M} _{\eta ''\rightarrow \eta } \approx  \mathcal{M} _{\eta '\rightarrow \eta } \times  \mathcal{M} _{\eta ''\rightarrow \eta '}$.%
\seqAAAaah{Leveraging the well-developed procedures for quantizing finite-dimensional systems}%
\seqAAAabe Once a collection of \emph{classical} partial theories has been put in the just mentioned factorized form, constructing a corresponding collection of \emph{quantum} partial theories is straightforward {\bseqJJJaat}. All we have to do is to quantize each $\mathcal{M} _{\eta }$ into a corresponding Hilbert space $\mathcal{H} _{\eta }$, and, since $\mathcal{M} _{\eta }$ is \emph{finite}-dimensional, we have at our disposal the full range of techniques known for quantizing systems with finitely many dof. From the usual rules of quantum mechanics we expect that, for any $\eta  \preccurlyeq  \eta '$, the factorization of phase spaces $\mathcal{M} _{\eta '} \approx  \mathcal{M} _{\eta '\rightarrow \eta } \times  \mathcal{M} _{\eta }$ will give rise to a tensor-product factorization of Hilbert spaces $\mathcal{H} _{\eta '} \approx  \mathcal{H} _{\eta '\rightarrow \eta } \otimes  \mathcal{H} _{\eta }$.%
\Mpar \vspace{\Sssaut}\hspace{\Alinea}In this way, we can obtain a collection of Hilbert spaces, with relations between them expressed in terms of tensor-products. But where is the promised quantum state space for the full field theory? The crucial realization, that goes back to Kijowski {\bseqJJJaaa}, is that we can now construct quantum states as collections $\big( \rho _{\eta } \big)_{\eta }$, where each $\rho _{\eta }$ is a density matrix (aka.~mixed or statistical state) on $\mathcal{H} _{\eta }$. In order for such a collection to describe a quantum state for the full theory, we need to ensure that the partial states $\rho _{\eta }$ are consistent with each other. Namely, whenever $\eta  \preccurlyeq  \eta '$, the states $\rho _{\eta }$ and $\rho _{\eta '}$ should lead to the same measurement probabilities for all dof.~retained by the coarser label $\eta $. Using well-known tools from statistical quantum mechanics, this means that we need to enforce \begin{equation}\label{parTraceSt}\rho _{\eta } = \mkop{Tr} _{\eta '\rightarrow \eta } \rho _{\eta '}\end{equation} where $\mkop{Tr} _{\eta '\rightarrow \eta }$ denotes the \emph{partial trace} on the tensor product factor $\mathcal{H} _{\eta '\rightarrow \eta }$.
Indeed, the partial trace satisfies the important property \begin{equation}\label{parTraceObs}\mkop{Tr}  \rho _{\eta } \,  \hat{f}_{\eta } = \mkop{Tr}  \big( \mkop{Tr} _{\eta '\rightarrow \eta } \rho _{\eta '} \big) \,  \hat{f}_{\eta } = \mkop{Tr}  \rho _{\eta '} \big( \mathds{1} _{\eta '\rightarrow \eta } \otimes  \hat{f}_{\eta } \big) = \mkop{Tr}  \rho _{\eta '} \,  \hat{f}_{\eta '}\end{equation} In order to be able to consistently impose eq.~(\ref{parTraceSt}), the way to reconstruct a coarser partial quantum state from a finer one should be unambiguous, in other words we should have $$\mkop{Tr} _{\eta ''\rightarrow \eta } = \mkop{Tr} _{\eta '\rightarrow \eta } \circ  \mkop{Tr} _{\eta ''\rightarrow \eta '}$$ for any three increasingly fine labels $\eta  \preccurlyeq  \eta ' \preccurlyeq  \eta ''$. Fortunately, this is automatically guaranteed by the quantum analogue of the classical consistency condition from {\seqBBBaaq} {\seqDDDabf}.
The quantum state space obtained in this way may be called \emph{projective}, since it arises mathematically as a projective limit (like the space $\mathcal{M} _{\text{lim}}$ mentioned in {\seqBBBaan}).%
\Mpar \vspace{\Sssaut}\hspace{\Alinea}Note that it is necessary in this construction to use density matrices rather than Hilbert space vectors (aka.~pure states). Indeed, the partial trace is the right tool to use when one is interested in the restriction of a quantum state to specific dof.~(thanks to the property stressed in eq.~(\ref{parTraceObs})) and it \emph{cannot} be defined as a map between pure states, because the partial trace of a pure state is often a mixed state and, reciprocally, the partial trace of a mixed state can appear pure\footnote{When $\rho _{\eta '}$ has the special form $\tilde{\rho } \otimes  \left| \psi  \middle\rangle\!\middle\langle \psi  \right|$, any statistical superpositions in $\tilde{\rho }$ will be traced out.}.
Another way of looking at this is that the distinction between pure states and mixed ones is always relative to a certain (sub)system, hence it is effectively useless in an infinite-dimensional theory: the only absolute notion of state purity would be the one defined with respect to the full theory, but such an absolute notion cannot have any experimental relevance, because it would require to measure quantum states against a \emph{complete} set of commuting observables, that is, to perform \emph{infinitely} many measurements.%
\Mpar \vspace{\Sssaut}\hspace{\Alinea}In {\seqBBBabg} more specific results have been obtained regarding the implementation of this general strategy. As is apparent from the discussion above, what we need is a consistent quantization scheme, namely a prescription to quantize each partial theory $\mathcal{M} _{\eta }$ in such a way that the Cartesian product factorizations, holding on the classical side, are indeed lifted to tensor-product factorizations on the quantum side, and that the quantization of observables agree, ie.~that we do have $\hat{f}_{\eta '} = \mathds{1} _{\eta '\rightarrow \eta } \otimes  \hat{f}_{\eta }$ as was assumed in eq.~(\ref{parTraceObs}).%
\Mpar \vspace{\Sssaut}\hspace{\Alinea}Geometric quantization {\bseqJJJaac} can be understood starting from the following observation. A very simple way of quantizing a finite dimensional phase space $\mathcal{M} _{\eta }$ would be to set $\mathcal{H} _{\eta } = L_{2}(\mathcal{M} _{\eta })$\footnote{Note that $\mathcal{M} _{\eta }$, as a phase space, admits a natural measure: the Liouville measure, which is singled out (up to a multiplicative constant) and has the important property of being preserved under any Hamiltonian flow, which ensures, in particular, that $\hat{f}_{\eta }$ will be self-adjoint whenever $f_{\eta }$ real.} and, for any observable $f_{\eta }$ on $\mathcal{M} _{\eta }$ $$\hat{f}_{\eta } \psi  =  -i \left\{ f_{\eta }, \psi  \right\}.$$ Then eq.~(\ref{quBrk}) is simply the Jacobi equation satisfied by the Poisson brackets. This naive Ansatz can be improved by incorporating some multiplicative action of $f_{\eta }$ in the definition of $\hat{f}_{\eta }$, yielding what is known as \emph{pre-quantization} \bseqHHHaac{section 8.2}.
Since the pre-quantization primarily depends on the Poisson bracket structure of the classical theory, and this structure is preserved by the projections $\pi _{\eta '\rightarrow \eta }$ as required in {\seqBBBaan}, it generically provides a consistent quantization scheme in the sense above {\seqDDDabh}.%
\Mpar \vspace{\Sssaut}\hspace{\Alinea}Pre-quantization is however not fully satisfactory (hence the 'pre' qualification) as it leads to Hilbert spaces that are simply too big.
Physically admissible Hilbert spaces are extracted by restricting ourselves to a subset of the pre-quantized Hilbert space \bseqHHHaac{chap.~9}: for example the usual \emph{position} representation of quantum mechanics is recovered by considering only the pre-quantum states $\psi (x,p)$ that are constant in $p$, ie.~only depend on $x$, while \emph{holomorphic} quantization is obtained by keeping only the $\psi $ that are holomorphic functions (with respect to a suitable complex structure on $\mathcal{M} _{\eta }$, eg.~$z = x+ip$). This extra step of the geometric quantization program is called imposing a \emph{polarization}. For the purpose of constructing projective quantum state spaces, the question is therefore whether polarizations can be imposed \emph{consistently} on each pre-quantized $\mathcal{H} _{\eta }$.%
\Mpar \vspace{\Sssaut}\hspace{\Alinea}Building on previous work by Okołów {\bseqJJJaas}, this question was examined in two concrete cases (namely, position quantization when the configuration spaces for all partial theories are Lie groups and the projective structure from {\seqBBBaan} is suitably compatible with the group structure, in {\seqBBBabi}, and holomorphic quantization when all projection maps $\pi _{\eta '\rightarrow \eta }$ are holomorphic, in {\seqBBBabj}) and the answer was found to be positive.
These results suggest that polarizations can be consistently imposed on the pre-quantized projective system whenever the classical precursors of these polarizations are compatible with the coarse-graining projections $\pi _{\eta '\rightarrow \eta }$.
Moreover, it may be possible to further generalize the quantization procedure by dispensing from the requirement that the classical factorizations hold \emph{globally} (ie.~dealing with a possibly non-trivial global topology, see {\seqBBBabk}).
\seqAAAabl%
\Mnomdefichier{ova3}%
\seqAAAabm{Relations to Standard Constructions: Improved Universality and Richer State Space}%
\seqAAAaaj{Embedding all vacuum sectors at once}%
\seqAAAabn Since the program exposed in the previous section is meant to improve on the vacuum-based quantization described in {\seqBBBaae}, we need to understand how the quantum state spaces obtained by both methods relate to each other. Suppose that we are given a vacuum state $\Omega $, which can be completely specified, as underlined in {\seqBBBaae}, by the expectation values it prescribes for any product of observables. Then, we can define a partial vacuum state $\Omega _{\eta }$, by simply forgetting about the expectation value of any product involving an observable that does not live on the partial theory labeled by $\eta $. Now, quantum theories describing finitely many dof.~do not suffer from the pathologies affecting quantum field theory: the states that can be written as density matrices on the Hilbert space $\mathcal{H} _{\eta }$ are likely to span a large part of the space of possible quantum states. Thus, it is not unreasonable to assume that each $\Omega _{\eta }$ corresponds to a density matrix $\rho _{\eta }$ on $\mathcal{H} _{\eta }$, and, since the $\Omega _{\eta }$ all stem from the same state $\Omega $ of the full theory, they are guaranteed to be consistent, ensuring that eq.~(\ref{parTraceSt}) holds. In other words, the vacuum $\Omega $ is likely to belong to the projective quantum state space constructed in {\seqBBBaah}, and with him all quantum states that only differ marginally from $\Omega $, ie.~the whole vacuum sector around $\Omega $ {\seqDDDabo}.
This reasoning suggests that the vacuum sectors corresponding to a large class of vacuums are included for free in the projective state space.%
\Mpar \vspace{\Sssaut}\hspace{\Alinea}In {\seqBBBabp}, we examined in particular the case where the $\rho _{\eta }$ obtained from $\Omega $ are actually \emph{pure} states for all $\eta $. Note that in contrast to the discussion so far, this is a very special situation, in the sense that pure states are fairly 'exceptional' in the space of all quantum states \bseqHHHaag{part III, theorem 2.2.18}; yet, it turns out to apply to the most interesting cases (such as the embedding of the usual Fock space, or of the \longAL state space mentioned in {\seqBBBaae}, into their projective counterparts, see {\seqBBBabq}).
Under this additional assumption, it is possible to prove that the embedding of the corresponding vacuum sector into the projective state space is injective (aka.~one-to-one) and its image can be precisely characterized.%
\Mpar \vspace{\Sssaut}\hspace{\Alinea}A different construction that could claim to include at once many vacuum sectors are the so-called infinite tensor products (ITP, {\bseqJJJaau}). These can be defined as huge direct sums of GNS Hilbert spaces $\mathcal{H} _{\Omega }$ (recall {\seqBBBaae}) for $\Omega $ running through a suitable class of vacuums. In other words, the different vacuum sectors are simply put \emph{side by side}.
Besides resulting in a large, non-separable Hilbert space, a drawback of this construction is that it is fairly fragile: it requires a decomposition of the dof.~into independent (commuting) subsets, with the slightest modification of this decomposition leading to \emph{inequivalent} ITP Hilbert spaces.
More seriously, it introduces spurious distinctions between physically indistinguishable states: since the different sectors appear as \emph{superselection sectors} (ie.~the observables are not sensible to quantum correlations \emph{between} these sectors), quantum superpositions of states in different sectors are indistinguishable from statistical superpositions.
By contrast, projective state spaces, while allowing for statistical superpositions of states in different vacuum sectors, do not produce such an unphysical redundancy.
This is exhibited by the fact that we can map the states on an ITP to states in a corresponding projective state space, in such a way that ITP states are mapped to the \emph{same} projective state whenever they are physically indistinguishable {\seqDDDabr}.%
\seqAAAabs{The case for countable collections of partial theories}%
\seqAAAabt While projective state spaces give access to a large class of states, they also have the advantage of providing a fairly explicit description of these states, in contrast to an approach where the quantum state space is directly defined as the set of \emph{all} possible quantum states (like in AQFT {\bseqJJJaai}, which we mentioned in {\seqBBBaae}).
In the latter case, quantum states have to be specified through the expectations values they prescribe for arbitrary products of observables\footnote{See however an alternative formulation, advocated in {\bseqJJJaav}, allowing to recast the full state space of AQFT in projective form.}.
The difficulty is that there are relations that these expectations values together must satisfy (in particular to ensure that all measurement probabilities will come out positive, see \bseqHHHaag{part III, def.~2.2.8}) making it non-trivial to actually construct valid states in this form and study their properties.
Arguably, quantum states expressed as collections of partial density matrices are also subjected to the satisfaction of extra relations (namely, the consistency condition from eq.~(\ref{parTraceSt})).
Yet, with an additional assumption on the \emph{size} of the label set, it is possible to obtain an explicit parametrization of all available projective states {\seqDDDabu}. %
\Mpar \vspace{\Sssaut}\hspace{\Alinea}Suppose first that our partial theories are labeled by an increasing sequence: say, for concreteness, $\big( \eta _{n} \big)_{n\geqslant 1}$, where the partial theory labeled by $\eta _{n}$ consists of lattice dof.~on a regular lattice with step $\nicefrac{a}{2^{n}}$. Then, the full quantum state space can be constructed very easily: indeed, each projective state can be constructed via a recursive process, starting from a density matrix $\rho _{{\eta _{1}}}$ on $\mathcal{H} _{{\eta _{1}}}$, and choosing, for any $n\geqslant 1$, $\rho _{{\eta _{n+1}}}$ in the pre-image $\mkop{Tr} _{{\eta _{n+1} \rightarrow  \eta _{n}}}^{-1} \left\langle  \rho _{{\eta _{n}}} \right\rangle $ (taking advantage of the fact that the partial trace is always surjective).%
\Mpar \vspace{\Sssaut}\hspace{\Alinea}The class of label sets for which a \emph{constructive} description of the quantum state space can be obtained in this manner is actually much broader. This follows from the following observation. If we have a projective state space build on a label set $\mathcal{L}$, then restricting it to a subset $\mathcal{K}$ does \emph{not} change the quantum state space, provided $\mathcal{K}$ is \emph{cofinal} in $\mathcal{L}$, ie.~that it satisfies the condition $$\forall  \eta  \in  \mathcal{L},\,  \exists  \kappa  \in  \mathcal{K} \mathrel{\big/}  \eta  \preccurlyeq  \kappa .$$ The projective state spaces are the same because coarser partial states can always be reconstructed from finer ones: if we know $\rho _{\kappa }$ for all $\kappa  \in  \mathcal{K}$, we can reconstruct any $\rho _{\eta }$ as $\mkop{Tr} _{{\kappa  \rightarrow  \eta }} \rho _{\kappa }$ for some $\kappa  \succcurlyeq  \eta $ {\seqDDDabv}.
For example, if we had chosen the sequence of lattice steps above to be $\nicefrac{a}{4^{k}}, k \geqslant  1$, we would in fact have been describing the \emph{same} state space, since $\left\{ 2k \middlewithspace| k \geqslant  1 \right\}$ is a cofinal part of $\left\{ n \middlewithspace| n \geqslant  1 \right\}$.%
\Mpar \vspace{\Sssaut}\hspace{\Alinea}Now, any \emph{countable} label set admits a cofinal sequence of increasingly fine labels (this follows easily from the directedness property mentioned in {\seqBBBaan}, see {\seqBBBabw}). Furthermore, the same observation also implies that it is not really the \emph{cardinality} of the label set that matters, but rather its \emph{cofinality} (the smallest possible cardinality of a cofinal part in it).
To summarize, constructive descriptions are readily available for all projective quantum state spaces built on label sets having countable cofinality.%
\seqAAAabx{Projective state spaces are robust with respect to the selection of partial theories}%
\seqAAAaby As each label extracts finitely many dof.~out of the full theory, expecting the label set to be of countable cofinality effectively requires that the full theory be spanned by countably many basic observables. This is a physically legitimate assumption, since it should be possible to determine and specify \emph{which} observable has been measured in a given experiment using only a finite amount of information.
This does not forbid the classical field theory to be formulated in terms of a continuum of dof.~for mathematical convenience, but we should keep in mind that infinitesimally close observables will be indistinguishable in practice, so that a dense and countable subset of basic observables should be sufficient to capture the entire physical content of the theory.%
\Mpar \vspace{\Sssaut}\hspace{\Alinea}Nevertheless, one might be worried about making the construction of the quantum state space dependent on the selection of a more or less arbitrary subset of observables. In particular, it seems that the \emph{universality} put forward in {\seqBBBaam} would suffer.
Continuing the example started in {\seqBBBabs} above, we could have chosen the sequence of lattice steps to be, say, $\nicefrac{a}{3^{p}}, p \geqslant  1$. In the absence of a background geometry, the choice of a sequence of finer and finer lattices would involve an even greater degree of arbitrariness, since there is no notion of 'regular lattices' in this case (see the construction in \bseqHHHaaw{section 3} and {\seqBBBabz}).
The natural question to ask is therefore whether these various options lead to dramatically different quantum state spaces.%
\Mpar \vspace{\Sssaut}\hspace{\Alinea}Remarkably, there is a simple condition that an increasing sequence of labels should satisfy to ensure that the projective state space built on it will be universal {\seqDDDaca}. Denote by $\mathcal{L}^{\extendedLabelSet }$ the complete set of all possible labels, among which a sequence has to be chosen. As explained above, if a cofinal sequence $\big( \kappa _{k} \big)_{k\geqslant 1}$ can be found in $\mathcal{L}^{\extendedLabelSet }$, the projective state space built on it coincides with the one that could be build on $\mathcal{L}^{\extendedLabelSet }$ and is thus universal. But if $\mathcal{L}^{\extendedLabelSet }$ is simply too big to admit such a cofinal sequence, we can relax the condition and only require the sequence $\big( \kappa _{k} \big)_{k\geqslant 1}$ to be \emph{quasi-cofinal} in the following sense: for any label $\eta $ in $\mathcal{L}^{\extendedLabelSet }$, it should be possible to slightly deform $\eta $ so as to make it a sublabel of one of the $\kappa _{k}$. More precisely, it should be possible {\seqDDDacb}:%
\seqAAAacc to let the deformation be arbitrarily small;%
\seqAAAacd and, if some of the dof.~contained in $\eta $ \emph{are} among the dof.~captured by the sequence $\big( \kappa _{k} \big)_{k\geqslant 1}$ (namely, if there exists a $\kappa _{{k_{o}}}$ that happens to be a sublabel of $\eta $), to demand that these particular dof.~be left untouched by the deformation {\seqDDDace}.%
\seqAAAacf In the case of 1-dimensional lattices (and by extension of square lattices in any dimension), it transpires from the analysis in {\seqBBBacg} that a sequence of finer and finer lattices will be quasi-cofinal if, and only if, the set of lattice nodes becomes dense as $k \rightarrow  \infty $.
This is in agreement with the physical discussion at the beginning of the present subsection: any dense subset of dof.~should be sufficient to capture the physical content of the theory.%
\seqAAAach Let us clarify what we precisely mean by saying that the projective state space on a quasi-cofinal sequence is 'universal' {\seqDDDaci}. If we have two different quasi-cofinal sequences, the set of dof.~captured by the first one will of course not be \emph{exactly} the same as the set of dof.~captured by the second one. However, an arbitrarily small deformation can map the first set into the second one, and, \emph{modulo} this approximate identification of the observables, the two quantum state spaces will coincide (ie.~any state in the first one will be in \emph{one-to-one} correspondence with a state in the second one, in such a way that the expectations values of the suitably identified observables match). An important corollary of this property {\seqDDDacj} is that symmetries of the original field theory can be represented as isomorphisms of the thus constructed quantum state space: indeed, a symmetry transformation, mapping a sequence $\big( \kappa _{k} \big)$ to a sequence $\big( \lambda _{l} \big)$, can be composed with a small deformation mapping $\big( \lambda _{l} \big)$ back to $\big( \kappa _{k} \big)$, to obtain a transformation that stabilizes the state space on $\big( \kappa _{k} \big)$.%
\Mpar \vspace{\Sssaut}\hspace{\Alinea}Note that this notion of universality should not be confused with the one expressed by Fell's theorem {\bseqJJJaax}. Fell's theorem is a result stating that the Hilbert space $\mathcal{H} _{\Omega }$ constructed from a vacuum state $\Omega $ as discussed in {\seqBBBaae} is approximately universal in the following sense: for any quantum state $\Omega '$ that can be built on the same algebra of observables as $\Omega $, any finite set of observables $f_{1},\dots ,f_{N}$ and any desired precision $\epsilon $, there exists a density matrix $\rho $ on $\mathcal{H} _{\Omega }$ such that the expectation values of these \emph{particular} observables, as computed with respect to $\Omega '$ vs.~$\rho $, will not differ by more than $\epsilon $. 
The most crucial difference between this result and the one discussed in the present subsection is that, if one want to fully take advantage of Fell's theorem to recover from $\mathcal{H} _{\Omega }$ a sufficiently universal quantum state space, one will effectively have to work, not in the space of density matrices on $\mathcal{H} _{\Omega }$, but in a suitable \emph{completion} thereof (namely, the completion with respect to the topology implied by the just spelled out statement of the theorem).
By contrast, a projective state space constructed on a quasi-cofinal sequence is already 'complete' enough: its sole limitation lies in the fact that the quantum states it describes \emph{a priori} only prescribe the expectations values of the observables in a certain countably-generated subset, safe in the knowledge that all other observables included in the (artificial but technically convenient) continuum classical field theory can be reconstructed from those at an arbitrary level of precision.
\seqAAAack%
\Mnomdefichier{ova4}%
\seqAAAaak{From Kinematical to Dynamical State Spaces}%
\seqAAAacl{The need for regularization}%
\seqAAAacm We have not, so far, addressed the question of how the equations of motion of the original field theory are to be implemented on a quantum state space constructed along the lines of {\seqBBBaam}.
This question raises two distinct kinds of issues: first, we need to understand how the dynamics can be expressed in a \emph{classical} theory which has been reformulated as a collection of partial theories as described in {\seqBBBaan}; second, we need to suitably quantize this classical dynamics.
The second part of the problem will involve the same tools as form usually part of a canonical quantization endeavor {\bseqJJJaay}, however the first part is specific to projective state spaces.
The core difficulty is that the dynamics is unlikely to respect the coarse graining introduced in {\seqBBBaan}. In a field theory we cannot, in general, write \emph{closed equations} involving only \emph{finitely} many degrees of freedom: the evolution of an infinite-dimensional system, when projected on a finite-dimensional truncation, is likely to be non-deterministic, because it is subject to back-reaction from all the discarded dof.~whose actual value is unknown.%
\Mpar \vspace{\Sssaut}\hspace{\Alinea}For the rest of the present section, we will adopt a parametrized description of the dynamics: by adding the time variable explicitly to the configuration space of our theory, solutions of the equations of motion can be represented as orbits or trajectories in the resulting extended phase space (see \bseqHHHaac{section 1.8} as well as the procedure illustrated in {\bseqJJJaaz}).
This way of looking at dynamics has the advantage that it generalizes seamlessly to gauge theories (where the solutions correspond to orbits under both time evolution and gauge transformations), and even to theories like general relativity, that do not refer to any external time variable (where time evolution is just another gauge transformation {\bseqJJJaba}).
In this language, the dynamics of the original field theory are said to be \emph{adapted} to a collection of partial theories $\big( \mathcal{M} _{\eta } \big)_{\eta }$ if they induce consistent partial dynamics on each $\mathcal{M} _{\eta }$ {\seqDDDacn}. Consistency here means that any orbit of the partial dynamics on a finer label $\eta '$ should project exactly to an orbit on a coarser label $\eta $. In this way, we can obtain projective families of orbits, which correspond precisely to the solutions of the full dynamics (left part of {\seqBBBaco}).
As stressed above, we cannot expect this ideal case to apply in any realistic field theory (unless we dispose \emph{beforehand} of an extensive understanding of the dynamics, that we can take into account when defining the collection of partial theories; this would suffer from the same limitations as underlined in {\seqBBBaae} in the context of vacuum-based quantization).%
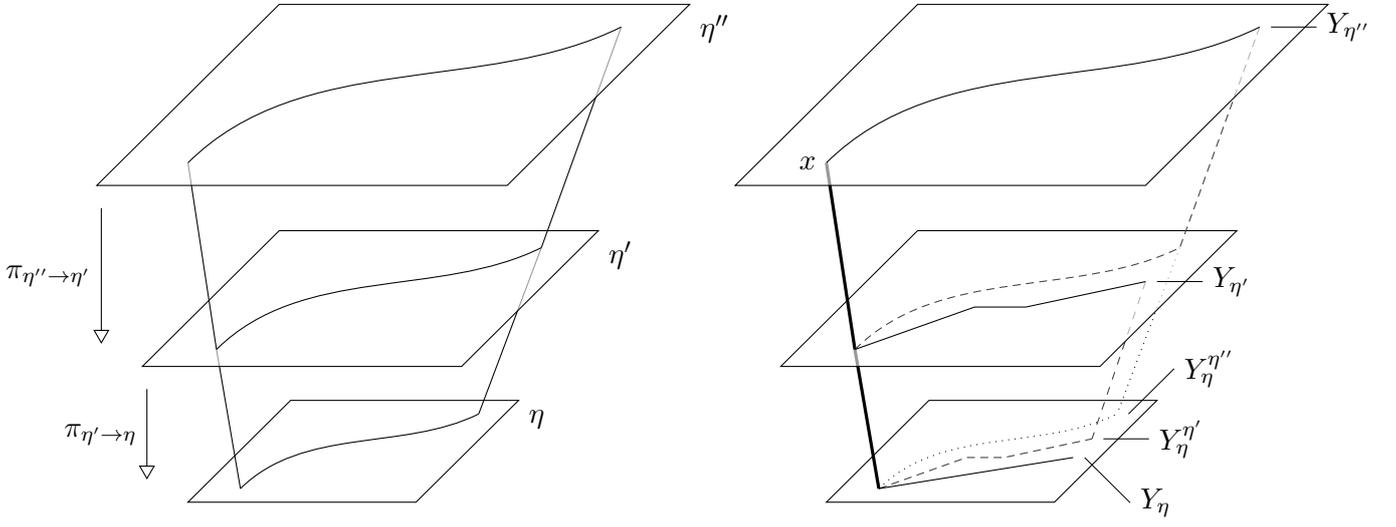
\seqAAAaco {Adapted (left side) versus regularized dynamics (right side)}{{\InputFigPacp}}%
\seqAAAacq What we can do instead {\seqDDDacr}, is to define, for each $\eta $, \emph{approximate} partial dynamics, that \emph{can} be formulated in closed form over $\mathcal{M} _{\eta }$ at the cost of not \emph{exactly} matching the desired full dynamics.\footnote{The approach taken in {\seqBBBacr} is a slight generalization of this proposal, in which additional parameters can be introduced to control the approximation. This is done by indexing the approximation scheme on a larger label set, namely a certain part of $\mathcal{L} \times  \mathcal{E}$, with the set $\mathcal{E}$ holding the extra approximation parameters.}
These approximate dynamics should become closer and closer to the exact one as we consider finer and finer labels $\eta $.
This implies that such a collection of approximate partial dynamics will \emph{not} be consistent in the sense above, because the partial dynamics on a finer label $\eta '$ are more accurate than the ones on a coarser label $\eta $.
Still, given some initial data $x(t=0)$ (or, more generally, given a point on a gauge fixing surface in $\mathcal{M} _{\infty }$ that intersects each orbit exactly once), we can consider the family of approximate orbits $\big( Y_{\eta } \big)_{\eta }$, where $Y_{\eta }$ denotes the orbit on $\mathcal{M} _{\eta }$ originating from (the projection of) $x$ (right part of {\seqBBBaco}).%
\Mpar \vspace{\Sssaut}\hspace{\Alinea}The idea behind this approach is to represent an exact solution of the full dynamics by the family of approximate solutions that converges to it.
Of course, not all families of successive approximations will converge to an exact solution: the initial data $x$ will in general have to satisfy suitable regularity conditions for the approximation scheme to converge.
But, importantly, the convergence of a given family can be studied completely within the projective setup, \emph{without} referring to the continuous theory $\mathcal{M} _{\infty }$. Let us define, for any $\eta  \preccurlyeq  \varepsilon $ $$Y^{\varepsilon }_{\eta } \mathrel{\mathop:}=  \pi _{\varepsilon \rightarrow \eta } \left\langle  Y_{\varepsilon } \right\rangle $$ ie.~the orbit at a finer level of approximation $\varepsilon $, projected back on $\mathcal{M} _{\eta }$. As stressed above the fact that the partial dynamics defined on $\mathcal{M} _{\eta }$ is only a coarse approximation is reflected by the deviation $Y_{\eta } \neq  Y^{\varepsilon }_{\eta }$. Convergence of the approximation is then expressed as the requirement that, for each $\eta $, the orbit $Y^{\varepsilon }_{\eta }$ converges, as $\varepsilon $ gets finer and finer, to (the projection of) an exact orbit $Y^{\infty }_{\eta }$. Now, the family $\big( Y^{\infty }_{\eta } \big)_{\eta }$ \emph{is} a projective family of orbits by construction, and as in the case of 'adapted dynamics' discussed above, it can be seen as the projective description of an exact solution of the full dynamics.%
\seqAAAacs{A framework for studying convergence at the quantum level}%
\seqAAAact The ability to discuss the convergence of the regularization scheme from within the projective setup is especially crucial when going over to the quantum theory, where there is no obvious equivalent of $\mathcal{M} _{\infty }$. Specifically, we will implement the dynamics on the quantum projective state space by turning the approximate partial dynamics on each $\mathcal{M} _{\eta }$ into a corresponding quantum dynamics on $\mathcal{H} _{\eta }$, constructing for any given initial data an associated family of approximate solutions, and study its convergence with the same strategy as on the classical side.
Note that a difference between classical and quantum dynamics is that solutions of the (approximate) quantum dynamics are just special quantum states, not \emph{subsets} of states as in the classical case: this effect can be observed already at the level of classical \emph{statistical} mechanics, where 'kinematical' states would be arbitrary probability distributions on the (extended) phase space, while solutions, or 'dynamical' states, would be those probability distributions that are \emph{constant} along the orbits.
Accordingly, a converging family of approximated solutions of the quantum dynamics will define a projective family of quantum states $\big( \rho ^{\infty }_{\eta } \big)_{\eta }$; in other words, such a family will provide a special state in the kinematical projective quantum state space.%
\Mpar \vspace{\Sssaut}\hspace{\Alinea}In {\seqBBBacu}, this procedure has been applied to the second quantization of the Schrödinger equation (aka.~non-relativistic quantum field theory {\bseqJJJabb}). One-particle quantum mechanics, seen as a classical field theory, can be quantized in a traditional way leading to a Fock space which is simply the Fock space describing an arbitrary number of non-interacting particles, each obeying the first-quantized theory.
Alternatively, a projective quantum state space can be built (the partial theories being obtained by orthogonally projecting the wavefunction on every finite dimensional subspace of the one-particle Hilbert space) and the dynamics can be regularized on it.
Interestingly, a domain of convergence for this regularization of the quantum dynamics can be delimited which is consistent with the traditional approach, in the sense that the dynamical projective quantum states it describes are precisely the ones in the Fock sector.
This sheds light as to why the Fock space is an appropriate arena for describing this \emph{non-interacting} theory and demonstrates how projective techniques can allow us to get a handle on the quantum dynamics without having to guess \emph{beforehand} a good vacuum: instead, a sector on which the dynamics can be well-defined should naturally emerge from the study of the convergence behavior of approximate solutions.
This also raises the intriguing open question of whether suitable domains of convergence will always take the form of a vacuum sector with respect to a certain (dynamics-aware) vacuum.
\seqAAAacv%
\Mnomdefichier{ova5}%
\seqAAAaag{Outlook: Bridging Canonical Quantization and Path-integral Approaches}%
\seqAAAacw Looking for a good quantum state space may seem like a preoccupation only relevant if we insist to rely on canonical quantization.
The argument here would be that we do not actually need any quantum state space: it is sufficient to know the quantum probabilities for the outcome of any arbitrary experimental protocol, and these probability distributions could themselves be reconstructed from their moments.
However, the so-called \emph{time-ordered} expectation values that can be computed from a path-integral formulation are \emph{not} the moments corresponding to the successive measurement of the corresponding observables, notwithstanding the fact that the measurement operations are, of course, causally ordered.
For example, to know the probability of measuring successively $f_{1} = a_{1}, \dots , f_{n} = a_{n}$, we need to compute\footnote{This is assuming the universe was initially in the vacuum state, but we can easily relax this questionable assumption by considering relative probabilities, eg. $\nicefrac{P(f_{1} = a_{1}, \dots , f_{n} = a_{n})}{P(f_{1} = a_{1})}$.} $$P(f_{1} = a_{1}, \dots , f_{n} = a_{n}) = \frac{\left\| \Pi _{1} \Omega  \right\|}{\left\| \Omega  \right\|} \frac{\left\| \Pi _{2} \Pi _{1} \Omega  \right\|}{\left\| \Pi _{1} \Omega  \right\|} \dots  \frac{\left\| \Pi _{n} \dots  \Pi _{1} \Omega  \right\|}{\left\| \Pi _{n-1} \dots  \Pi _{1} \Omega  \right\|} = \sqrt{\left\langle  \Omega  \middlewithspace| \Pi _{1} \dots  \Pi _{n} \Pi _{n} \dots  \Pi _{1} \middlewithspace| \Omega  \right\rangle }$$ where $\Pi _{i}$ denotes the spectral projector on the eigenvalue $a_{i}$ of $\hat{f}_{i}$.
Knowing all time-ordered expectation values is thus \emph{a priori} not sufficient to recover the physically interesting moments (note the mirroring in the formula above: from right to left, the observables appear first in causal-order and are then repeated in \emph{reverse} causal order). Reciprocally, it turns out that if we had the vacuum-vacuum expectation values for product of operators in arbitrary orders, which would encode all the physically interesting information, we could in fact also reconstruct the quantum state space, via the GNS construction mentioned in {\seqBBBaae} (in this context also known as Wightman reconstruction {\bseqJJJabc}).%
\Mpar \vspace{\Sssaut}\hspace{\Alinea}Now, on Minkowski spacetime, the expectation values for products of observables in arbitrary order can \emph{all} be reconstructed from the time-ordered ones (or, equivalently, the quantum expectation values for the different orderings of operators can be all reconstructed via analytic expansion from the \emph{Euclidean} path-integral: different orderings then arise as different prescriptions for going around singularities, as was proved by Osterwalder and Schrader {\bseqJJJabd}).
It is however not clear how the fairly subtle complex analysis underlying this result could be generalized to non-static background spacetimes (although results exist when a timelike Killing vector is available {\bseqJJJabe}), or even to background-independent theories.
This is the situation where disposing of a (kinematical) quantum state space can be valuable even for the path-integral approach.
We can then, instead of only computing vacuum-vacuum expectation values, develop the path-integral between \emph{arbitrary} boundary quantum states (like is done eg.~in the spin foam approach {\bseqJJJabf} to quantum gravity, leaning against the \longAL Hilbert space mentioned in {\seqBBBaae}) and this is indeed an alternative way of capturing all the physically interesting information (since we can then insert a decomposition of the identity in the middle of the above formula).%
\Mpar \vspace{\Sssaut}\hspace{\Alinea}Projective quantum state spaces seem particularly well-suited to play the role of supportive state spaces for boundary-aware path-integral formalisms. On one hand, they implement the same kind of coarse-graining/refining which is used in practice to compute, through a limiting process, path-integral expressions. On the other hand, they anticipate the loss of privileged status for the vacuum state that is the natural consequence of allowing non-vacuum boundaries.
Reciprocally, techniques that have been developed in the context of path-integral approaches could be applied to the strategy presented in {\seqBBBaak}, to help with the study of \emph{convergence}, and with the inclusion of \emph{renormalization} (which was not needed for the simple example considered in {\seqBBBacu} but is likely to come into play when considering interacting theories).%
\seqAAAacx \seqAAAacy
\seqAAAacz%
\Mfin%
\Mybibliography{overa}%
\end{document}